# TranspoGene and microTranspoGene: transposed elements influence on the transcriptome of seven vertebrates and invertebrates

Asaf Levy, Noa Sela and Gil Ast*

Department of Molecular Genetics and Biochemistry, Tel-Aviv University Medical School, Tel Aviv 69978, Israel

**ABSTRACT**

**Transposed elements (TEs) are mobile genetic sequences. During the evolution of eukaryotes TEs were inserted into active protein-coding genes, affecting gene structure, expression and splicing patterns, and protein sequences. Genomic insertions of TEs also led to creation and expression of new functional non-coding RNAs such as microRNAs. We have constructed the TranspoGene database, which covers TEs located inside protein-coding genes of seven species: human, mouse, chicken, zebrafish, fruit fly, nematode and sea squirt. TEs were classified according to location within the gene: proximal promoter TEs, exonized TEs (insertion within an intron that led to exon creation), exonic TEs (insertion into an existing exon) or intronic TEs. TranspoGene contains information regarding specific type and family of the TEs, genomic and mRNA location, sequence, supporting transcript accession and alignment to the TE consensus sequence. The database also contains host gene specific data: gene name, genomic location, Swiss-Prot and RefSeq accessions, diseases associated with the gene and splicing pattern. In addition, we created microTranspoGene: a database of human, mouse, zebrafish and nematode TE-derived microRNAs. The TranspoGene and microTranspoGene databases can be used by researchers interested in the effect of TE insertion on the eukaryotic transcriptome. Publicly available query interfaces to TranspoGene and microTranspoGene are available at http://transpogene.tau.ac.il/ and http://microtranspogene.tau.ac.il, respectively. The entire database can be downloaded as flat files.**

## INTRODUCTION

Transposed elements (TEs) are mobile genetic sequences that constitute 45%, 38%, 15–22%, 12%, and 9% of the human, mouse, fruit fly, nematode and chicken genomes, respectively (1–4). TEs are distinguished by their mode of propagation: short interspersed repeat elements (SINE), long interspersed repeat elements (LINE) and retrovirus-like elements with long-terminal repeats (LTR) are propagated by reverse transcription of an RNA intermediate. In contrast, DNA transposons move through a direct 'cut-and-paste' mechanism (5).

TEs have shaped the eukaryotic genome and transcriptome in several ways. Mammalian evolution was notably affected by TEs through their contribution to genetic diversity, genomic expansion, genomic content and genomic rearrangements (5,6). Although introns comprise only 24% of the human genome, 60% of all TEs are located in introns (7). L1 elements can significantly decrease mammalian gene expression when inserted within introns, due to inadequate transcriptional elongation (8). Intronic TEs from all human and mouse TE families can gain mutations leading to creation of an additional exon in a process known as 'exonization'. The most prominent TE involved in exonization process is the primate-specific Alu (7,9,10). TEs can also insert into exons, a phenomenon that is common mainly in the 5′ and 3′ untranslated region (UTR) exons, at least in human and mouse (7). TE insertion elongates the UTR and can serve as new cis-acting element such as microRNA-binding site (11).

TE insertion can directly affect gene expression when the TE is incorporated into a gene promoter. The TE sequence may provide transcription factor binding sites previously not present in the promoter (12). About 25% of human promoter regions contain TE-derived sequences, including many experimentally characterized cis-regulatory elements (13). For example, some functional binding sites of the transcriptional repressor REST originated from duplication and insertion of RE1-containing L2 and Alu retrotransposons (14). Not only protein-coding genes are affected by TE insertion. A recent work by Borchert et al. (15) demonstrated that several human microRNAs are transcribed by RNA polymerase III through promoters and/or terminators derived from the Alu retrotransposon. TEs can also control eukaryotic genes epigenetically when inserted within or very close to these genes (16,17).

TE insertion may create new functional non-coding RNAs (ncRNAs). The BC200 gene is a neuronal, primate-specific ncRNA derived from a FLAM-C like monomeric Alu (18). Smalheiser and Torvik (19) showed that a few mammalian microRNA precursors are derived from intronic insertion of two adjacent LINE retrotransposons in opposite orientation, creating a hairpin structure that serves as microRNA precursor. Other TEs, such as the human MADE1, have an intrinsic hairpin structure and serve as microRNA precursors when inserted into transcriptionally active genomic regions (20). Many of the newly identified Piwi-interacting RNAs (piRNAs) are derived from transposons and play a role in transposon silencing in zebrafish germ cells (21).

TE-related rearrangement events may be deleterious in terms of fitness of the organism. More than 25 human genetic diseases are attributed to TE-related rearrangements (22,23). Accumulation of TEs is also associated with decrease in fitness in fruit fly (24). TE action contributes to fitness in other cases, however. For example, it seems that a LINE-1 retrotransposon catalyzed the creation of a chimeric gene that confers resistance to HIV-1 infection in owl monkeys (25).

We previously generated a database of Alu elements incorporated within protein-coding genes (26). Other groups created other database specializing in LINEs (27) and HERVs (28). All these databases focused on the insertion of typical TE family members exclusively in the human genome. The ScrapYard database (29) contains TEs in a few vertebrate genomes but it is based on Genbank only and it was updated in January 2002 for the last time. We have now created TranspoGene: a database unifying all existing TE families for seven vertebrate and invertebrate species, so that researchers interested in the influence of various TEs on a specific gene can find this data in one location. The TranspoGene website also includes another unique database: microTranspoGene. This database contains human, mouse, zebrafish and nematode microRNA derived from TEs either transcriptionally (from RNA pol III promoters/terminators in Alu retrotransposons) or structurally (through creation of the microRNA precursor hairpin structure by TE insertion). The TranspoGene database and website allow the user to retrieve important data such as the location and sequence of the TE with respect to the gene and the genome, supporting transcript evidence, annotation of the gene and protein (including related diseases), extensive splicing data for exonized TEs, and the evolutionary change undergone from the particular TE's consensus sequence.

## MATERIALS AND METHODS

### Dataset of exonized TEs and intronic TEs in protein-coding genes

Human NCBI 35 (hg17, May 2004), mouse NCBI33m (mm6, March 2005), chicken (galGal3, May 2006), zebrafish (danRer4, March 2006), fruit fly (dm2, April 2004), *Caenorhabditis elegans* (ce2, March 2004) and sea squirt (ci2, March 2005) assemblies were downloaded, along with their annotations, from the UCSC genome browser database (30). EST and cDNA mapping were obtained from chrN_intronEST and chrN_mrna tables, respectively. TE mapping data were obtained from chrN_rmsk tables and TE sequences were retrieved from appropriate genomes using the mapping data. A TE was considered to be intragenic if there was no overlap with ESTs or cDNA alignments; it was considered intronic if it was found within an alignment of an EST or cDNA in an intronic region. Finally, a TE was considered as an exonized TE if it was found within an exonic part of the EST or cDNA, if it possessed canonical splice sites, and if it was not the first or last exon of the EST/cDNA. Next we associated the intronic and exonized TEs with genomic positions of protein-coding genes. For human and mouse genomes we used the list of known genes from UCSC table browser after removal of protein-coding genes with relatively low reliability (received from Fan Hsu, personal communication). These removed genes had no corresponding Protein Data Bank, RefSeq mRNA or Swiss-Prot protein entries. For the other five genomes, which are less completely annotated than human and mouse genomes, we used RefSeq (31) genes tables from the UCSC table browser (30). Positions of the TE hosting intron/exon and the mature mRNA were calculated using the gene tables mentioned above. Association of the gene to the mRNA and protein accessions and description from RefSeq and Swiss-Prot was done through the kgXref and refLink tables in the UCSC genome browser database (30).

### Relevant data for exonization events

The 5′ and 3′ splice sites sequences of each TE-derived exon were retrieved from the genome. Inclusion level of the exon was calculated by dividing the number of deposited transcripts supporting exon inclusion by the total number of transcripts containing this exon. The splice mode (exon skipping, constitutive splicing, alternative 5′/3′ splicing) involving the exon was determined based on EST and cDNA data. Expression of human exonized TEs was based on expression in 11 normal human tissues determined using the Affymetrix exon chip 1.0; the AffyHumanExon table was obtained from the UCSC genome browser database (30). Data was associated with TE according to genomic position overlap. A tissue was associated with a specific exonized TE if the log-ratio of median normalized signal in this tissue was $\geq 0.75$.

### Dataset of TEs insertions within exons

Using Galaxy (32) we downloaded from the UCSC table browser (33) the Repeat Masker (rmsk)

tables for the seven organisms of interest. We also downloaded the exons of the human and mouse filtered genes (see above) and the exons of the RefSeq genes (31) for the rest of the organisms. We searched for intersection between the genomic intervals of TEs in rmsk (repeat classes: DNA, LTR, SINE and LINE) and the exons, where the TE was fully contained within an exon.

### Dataset of TEs in putative proximal promoters

Transcription start sites (TSS) were retrieved from the filtered known gene tables of human and mouse genomes (see description above) and the RefSeq (31) gene tables for the five other organisms. In some cases, more than one TSS existed per gene. Using the UCSC table browser (33), a region of 250 bp upstream to the TSS was obtained. TEs were defined as DNA, LTR, SINE and LINE classes and were obtained from the rmsk table of each of the seven species in Ref. (33). Association between the TEs and putative proximal promoter was based on genomic position overlap.

### Dataset of microRNA precursors derived from TEs

We downloaded the positions of human, mouse, chicken, zebrafish, fruit fly and nematode microRNA precursors from release 10.0 of miRBase (34). We searched using Galaxy (32) for position intersection of microRNA precursors with TEs (repeat classes DNA, LTR, SINE and LINE) from rmsk table of each genome taken from the UCSC table browser (33). Human microRNAs with significant experimental support for expression from RNA pol III Alu promoters were received from Glen Borchert from Iowa University (personal communication) based on Ref. (15).

### Human disease data

All human genes in the tables of TranspoGene were associated with diseases data from the OMIM database (35). The OMIM Morbidmap table was downloaded. The filtered known human genes table (see above) was associated with diseases from Morbidmap table through gene aliases taken from kgAlias table from the UCSC table browser (33).

### Alignments to TE consensus sequence

The TEs in TranspoGene were aligned to the consensus sequences using Repeat Masker (www.repeatmasker.org) version open-3.1.6, default mode. Repeat Masker was run with cross_match version 0.990329, RepBase Update 9.11 (36), RM database version 20050112.

### Modifying genomic positions to updated genome assemblies

We converted TranspoGene genomic positions of TEs and their associated genetic elements using the liftOver tool obtained from the UCSC genome browser (30). The following conversions were done: human: hg17 → hg18, mouse: mm6 → mm9, fruit fly: dm2 → dm3, nematode: ce2 → ce4.

### Database and website

The TranspoGene data is stored in a relational database, implemented with MySQL. The TranspoGene interface was created using HTML and Java script and allows access to database through HTML form. CGI script written in Perl dynamically translates the data entered by the user into the appropriate query to the database. The entire database can be downloaded as flat files from the download section.

## RESULTS AND DISCUSSION

Until a few years ago TEs were often referred to as 'junk DNA'. Makalowski suggested that TEs should be viewed as genomic symbionts that create a 'genomic scrap yard', the source of 'junk' that natural selection utilizes in its evolutionary experiments (37). Recent studies revealed the important role of these mobile elements in the evolution of eukaryotic genomes, the regulation of cellular processes, and their involvement in diseases. We have thus constructed a database of TEs from seven organisms to aid in effort to analyze the influence of TEs on the transcriptome of eukaryotes. Since construction of a totally complete database would be virtually impossible, we focused on data of high quality. This was done in several ways:

(i) We supplied transcript accessions (EST/cDNA) to support inclusion and skipping of exonized TEs and intronic TEs, respectively.
(ii) Human and mouse genomes are annotated with genes of low protein-coding potential. For example, genes of millions nucleotides long, supported by very few transcripts that often overlap with annotated genes. Therefore, we used a list of genes for which there is strong evidence for protein production: genes with corresponding entries in PDB or Swiss-Prot or corresponding RefSeq mRNA. These human and mouse genes lists are filtered lists and shorter than the whole UCSC known genes lists used by Sela *et al.* (7). Therefore, the numbers of exonized TEs, intronic and exonic TEs are different in TranspoGene database and in Sela *et al.* research. For the other five organisms we used the high-standard RefSeq genes (31).
(iii) We improved the experimental support for exonizations, which were originally based on ESTs. For human exonized TEs, we added expression data in healthy tissues from Affymetrix exon chips (38). There were 414 (29%) exonized TEs of those with EST evidence expressed in at least one of the 11 human healthy tissues hybridized on the microarray. The rest of the exonized TEs either did not have a designed probe on the chip (31%) or might be expressed in tissues that were not examined (40%). Consisting with our previous publications, most of TE exonizations are alternatively spliced (7,9,10,39). We calculated the inclusion level of the exon from the number of deposited transcripts supporting exon inclusion divided by the total number of transcripts containing the flanking exons. We annotated the exonized TE with the splice mode involved: exon skipping, constitutive splicing, alternative 5′/3′ splicing. We also supplied the splice site sequence for each exonized TE.

**Table 1.** TranspoGene and microTranspoGene content for human and mouse genomes

|  | Exonized TEs | Intronic TEs | Exonic TEs | TEs in putative promoters | Protein-coding genes containing any TE (percentage out of the total number of genes) | microRNA genes expressed from RNA pol III Alu promoter or terminator | microRNA precursors structurally derived from TE |
|---|---|---|---|---|---|---|---|
| Human | 1417 | 1 260 516 | 6878 | 2021 | 14 783 (88.8%) | 14 | 68 |
| Mouse | 342 | 652 218 | 5326 | 1058 | 12 027 (87.2%) | 0 | 17 |

(iv) To avoid false positives in the identification of TEs located in putative promoter regions, we used only predicted proximal promoters, which are located just upstream (within 250 nt of the gene's TSS.

(v) The microTranspoGene database is based on published data, some of which is experimentally validated (15,19,20).

The TranspoGene and microTranspoGene databases contain millions of TEs with validated or predicted effect on the transcriptome of seven species. Table 1 exemplifies TranspoGene and microTranspoGene content for human and mouse genomes, both of which share a very high TE content and many annotated protein-coding genes.

### Website usage

The TranspoGene website is user-friendly and allows easy searching of the database. Through the TranspoGene webpage (available at http://transpogene.tau.ac.il) the user initially selects genomic area of interest: gene name, protein/mRNA accession from RefSeq or Swiss-Prot, or absolute genomic positions in a specific organism. The user can also insert a list of gene names, possibly from different organisms. The MySQL wildcard sign (%) can be also used for searching genes with similar names. Afterwards the user selects the TE family of interest, where multiple selection of TE families within an organism and/or between organisms is enabled. Finally the TE type according to location in the gene is selected: intronic TE, exonized TE, TE in proximal promoter or exonic TE (multiple types can be selected). The result of the query is provided both in a table format and in a file ready for download. An example for the result of TranspoGene search is given in Figure 1. The results contain links to RefSeq (31), Swiss-Prot (40), OMIM (35) and to the UCSC genome browser (41) in order to provide further information about the relevant transcript, protein, disease and genomic region, respectively. To detect mutations that accumulated in a given TE, the user can view the alignment file of the TE sequence to its consensus sequence, through a link given in the query results table. For exonized TEs, the overlap between the exon and its overlapping TE is emphasized by use of different colors. The microTranspoGene webpage (http://microtranspogene.tau.ac.il/) allows selection of either microRNA precursors structurally derived from TEs or human microRNA genes transcribed by RNA polymerase III using Alu promoters/terminators. It allows searching for a specific microRNA precursor in the database, selection of all relevant microRNAs of a specific species or all microRNA precursors in the database from all organisms. Supplementary Figure 1 demonstrates partial microTranspoGene results for searching human microRNA precursors structurally derived from TEs. The entire database can be downloaded as flat files from the download link.

### SUPPLEMENTARY DATA

Supplementary Data are available at NAR Online.

### ACKNOWLEDGEMENTS

This work was supported by the Cooperation Program in Cancer Research of the Deutsches Krebsforschungszentrum (DKFZ) and Israel's Ministry of Science and Technology (MOST), by a grant from the Israel Science Foundation (1449/04 and 40/05), GIF, ICA through the Ber-Lehmsdorf Memorial Fund and DIP. N.S. is funded in part by EURASNET. Fan Hsu from the UCSC genome browser team kindly supplied us the high-quality known genes tables of human and mouse. Glen Borchert and Beverly Davidson from Iowa University kindly supplied us with part of the data used for microTranspoGene. We would like to thank Amiel Dror for his help in graphical design. We also wish to thank Itay Menachem, the head of the Life Sciences IT Unit, and the Unix system specialist Hila Afargan, for their assistance in the infrastructure design and implementation of the needed system that enables us to run TranspoGene. Funding to pay the Open Access publication charges for this article was provided by EURASNET.

*Conflict of interest statement*. None declared.

| TE family | TE name | Chromosome | TE strand | Repeat positions | Example for transcript including the exon | Transcript strand | Exon start | Exon end | 3' splice-site |
|---|---|---|---|---|---|---|---|---|---|
| human_Alu | AluJo | chrX | - | 19124096-19124374 | BP252370 | + | 19124285 | 19124398 | tactag |

| Exon sequence: overlapping part with TE seq is in red, the rest is in black | 5' splice-site | TE sequence over transcript: overlapping part with exon is in black, the rest is in red | TE orientation to gene | Gene region |
|---|---|---|---|---|
| AGACATGGTCTTGCTACGTTGCCCAGTC TGGTCTCCATCTCCAGGCTCAAGCAGTC CTCCCACCTCGGCCTCCCAAAGTGCTG GGATTACTCTCACTCTCTTAAAACCAGG CAG | CAGgtaggg | GGACAGCTCACTGCAGCCTTAGCCTGCTG GGCTCAAGCGATCCTCCTGCCTTAGCCTC CTGAGTAGCTGGGAACACAGGCATGTGCC ACCACCACACCCAGCCAATTAAAAAAATTT TTTTTTTACTAGAGACATGGTCTTGCTACGT TGCCCAGTCTGGTCTCCATCTCCAGGCTCA | antisenese | CDS |

| Number of transcripts holding exon | Number of transcripts skipping exon | Inclusion level | Splice mode | Organism | Gene name | Gene strand | Refseq accession protein | Protein description |
|---|---|---|---|---|---|---|---|---|
| 17 | 290 | 0.0553745928338762 | exon skipping | human | NM_000284 | + | NP_000275 | pyruvate dehydrogenase (lipoamide) alpha 1 |

| Gene related diseases | Expression in normal tissues in Affymetrix exon chip | Location of TE on refseq mRNA | Alignment to concensus sequence (press the link) |
|---|---|---|---|
| Leigh syndrome, X-linked; Pyruvate dehydrogenase deficiency | Heart | intron 1/10 | 17251 |

**Figure 1.** The TranspoGene results table generated after a search for human Alu exonizations in the RefSeq mRNA accession NM_000284. The original single result row was spilt into four vertical rows for a better view.